# Testing the influence of the temperature, RH and filler type and content on the Universal Power Law for new reduced graphene oxide TPU composites.

**J Gómez[1*], E Villaro[2,3], A Navas[2,4] and I Recio[1,4]**

[1] Avanzare Innovacion Tecnologica S.L. Avda. Lentiscares 4-6 26370, Navarrete, Spain.
[2] Instituto de Tecnologías Químicas de La Rioja (Inter-Química), San Francisco 11, 26370, Navarrete, Spain.
[3] Departamento de Química Inorgánica y Técnica, Facultad de Ciencias UNED, 28040 Madrid, Spain.
[4] Departamento de Química – Centro de Investigación en Síntesis Química. Universidad de La Rioja, 26006, Logroño, Spain

julio@avanzare.es
evillaro@interquimica.org
anavas@interquimica.org
irecio@avanzare.es

Julio Gomez Cordon
julio@avanzare.es

**ABSTRACT**

In this paper, 6 different reduced graphene oxide (rGO) were prepared by a modified Hummers' method and reduced by thermochemical methods. rGO materials were intentionally prepared to obtain different BET and thickness and oxygen content maintaining constant the lateral size to compare its performance on thermoplastic polyurethane (TPU) matrix. Microstructure and the effect of the incorporation of rGO on the hardness and electrical properties of TPU were investigated. It has been studied the temperature and humidity dependence of the electrical conductivity and the sensitivity and the response time to humidity changes have been determined. Influence of the filler content, temperature and humidity on the Jonscher's universal power law (UPL) for ac conductivity vs frequency and its fitting parameters A and n were determined. It has been observed an anomalous behaviour (according to UPL) and a linear correlation between log A and n independently of the filler content, humidity and temperature, however there is an influence of the rGO used for the preparation of the composite. To study the transport mechanisms the experimental results were adjusted to the equation $\sigma = \sigma_0 \exp[-(T_{Mott}/T)^\gamma]$ and the maximum adjustment for $\gamma = 1/4$ like other carbon nanocomposites however there is not an unequivocal behaviour.

## 1. Introduction

Graphene based nanocomposites have attracted much attention in the last decade to improve physicochemical properties of polymer matrices: electrical and thermal conductivity, mechanical performance [1-6].

Thanks to the particular chemical structure of thermoplastic polyurethanes (TPUs), characterized by hard and soft segment, these materials are very versatile offering a wide range of service temperatures and hardness options and excellent chemical and mechanical performance [7]. It is possible to increase and adapt mechanical or electrical properties of TPU resins using nanofillers, such as graphene in the preparation of composites. These properties make TPUs suitable for their use in several products and different industries such as footwear, engineering, building & construction, automotive, hose & tubing, wires & cables, and medical, and growth is expected in automotive, engineering and medical applications [8].

Currently, a variety of techniques have been developed to prepare graphene nanosheets (GNS). Reduced graphene oxide (rGO) has attracted much attention to obtain large lateral-size graphene materials, due to of their excellent electrical conductivity properties, and scalability, that makes it in an alternative to other carbon-based materials such as CNTs, Carbon Blacks and conductive graphite materials.

Among chemical oxidation methods, Hummers' method [9] has become the most popular approach to obtain graphene oxide as starting material in large scale production of graphene sheets by reduction methods: chemical or thermal. However, graphene material produced by this method contains a high amount of $sp^3$ defects. Chemical reduction and thermal annealing can be used for reducing $sp^3$ defects. Beside thermal, chemical and thermochemical reduction of graphene and graphite oxides are easy customizable and versatile methods for preparing different types of reduced graphene oxide grades. [10-12]

We have prepared and characterized 6 different rGO materials, controlling the exfoliation step. We have achieved low oxygen content (2,6%) for RGO1 and RGO2, similar lateral size, while the specific surface area (SSA), measured by BET isotherm, has shown significant differences in average thickness. We have also modified the reduction step to obtain rGO materials with different oxygen content and defects concentration. We produced rGO-TPU composites by solution blending with these rGO.

We published the preparation and measurement of the electrical conductivity of TPU-rGO nanocomposites using 20x20 μm graphene sheets, which shows low percolation limit [13].

Recently, it has been reported high electrical conductivity over 10 S/m at 10% of loading in graphene nanoplatelets/CNTs/PU composites and similar electrical conductivity ($10^{-4}$ S/m) at 1% of loading than the composites prepared for this study that show similar conductivity than previously reported ones [14-17]. Easy and quick preparation of these composites is one of their main advantages; however, we also want to highlight their high sensibility to humidity.

Empirical Jonscher´s UPL allows studying conduction mechanism into disordered matrices. The measured ac conductivity $\sigma(\omega)$ of conducting and semi-conducting materials is characterized by the transition above a critical (angular) frequency $\omega_0$ from a low-frequency dc plateau to a dispersive high-frequency region [18-25]. Anomalous power law dispersion has been observed in all kind of materials: single crystals, polymers, glasses, technical ceramics, conductive polymer composites, etc. [26-28]. Mauritz has recently reported a linear evolution of log A *vs* n and has opened the question of the link between A and n and the influence of the different materials [19]. In this paper, we test UPL behaviour of the prepared rGO/TPU composites analysing the relationship between log A *vs* n. We have observed linear evolution log A *vs* n at different filler contents, temperatures and humidity levels and there is a different evolution depending on the rGO used in the production of the nanocomposite.

In this paper, we also explore the effects of the average thickness of rGO in electrical properties of TPU composites. Temperature and humidity effect in the electrical conductivity are also analysed in addition to their potential application in low cost humidity self-sensing material.

## 2. Experimental

*2.1. Materials*

Flake Graphite powder with particle size of 600 μm was obtained from Grafitos BARCO, Spain. For preparation of rGO samples; $KMnO_4$, concentrated $H_2SO_4$, sodium hydroxide, hydrochloric acid and ascorbic acid were bought to COFARCAS (Spain). Thermoplastic polyurethane (TPU) GOLDENPLAST049 A85 NP30 with a density of 1,16 g/cm$^3$ (at 20ºC) was used as received. *N,N*-dimethylformamide 99,9% (DMF; Labkem), methanol, and isopropanol (Cofarcas Spain), were used as received for the study.

*2.2 Synthesis of graphene sheets*

The 2 rGO were prepared by a modified Hummers' method using flake graphite powders as the starting material.

*rGO Preparation:*

Graphene oxide was prepared using a modified Hummers' method [9] in $H_2SO_4$. Starting from graphite flakes (20 g) and using a proportion of graphite/$KMnO_4$ 1:3,75. The reaction temperature inside the reactor was kept between 0 and 4 ºC during the oxidants addition (48 h). After that time, resulting solution was slowly warmed up to 20ºC and maintained for 72 hours of reaction. To remove the excess of $MnO_4^-$, $H_2O_2$ solution was added to the reaction mixture and stirred overnight. After sedimentation, the solution was washed with a mechanical stirred HCl 4 %wt solution by 2 h. The solid was filtered off, diluted in osmotic water (4 %wt based on dry GOx) and stirred in a Dispermat LC75 using a cowless helix for 15 min at 1000 rpms and at 15000 rpms for 5 minutes. This solution was sonicated with a HIELCHER UP400S using a H22 sonotrode at maximum amplitude. For rGO1, the time of ultrasonication was 20 minutes and 45 minutes for rGO2, rGO3, rGO4, rGO5, rGO6. Temperature during sonication was controlled using a cooling bath and it was kept below 70 ºC. To the GO dispersion (10 g in 1,75 litres of water) ascorbic acid (8,75 g) was added and the mixture was refluxed overnight at atmospheric pressure. The solid was filtered off and air-dried. For the preparation of the thermochemically reduced rGO, the chemically reduced GO, was placed in an oven under Ar atmosphere for 20 min at: 1000 ºC (rGO1 and rGO2), at 1200ºC (rGO3), 900ºC (rGO4), 700ºC (rGO5), 200ºC (rGO6) . rGOs were obtained as a black solid with 0,004 (rGO1) and 0,002 g/ml (rGO2 to rGO6) of apparent density.

*2.3. TPU/rGO nanocomposites preparation*

rGO-TPU nanocomposites (from 0.25% to 1%wt) were prepared by solution blending. TPU was dissolved in DMF, and the required quantity of rGO was dispersed in DMF and sonicated in an ultrasound bath at 40 KHz. TPU solution was added over the rGO dispersion and the mixture was mechanically stirred at 1000 rpm in a DISPERMAT LC75 using a cowless helix of 50 mm diameter in order to get a homogeneous solution. After that, it was poured down slowly into methanol to precipitate the rGO-TPU nanocomposite. The material obtained was dried overnight at 70°C in a vacuum oven.

*2.4. Sample preparation and characterization*

For microscopy characterization (SEM or TEM), the samples of rGO were dispersed in isopropyl alcohol and sonicated with a Hielscher UP200S sonicator for 15 minutes follow by 20 minutes in a COBOS bath sonicator.

The morphology of rGO and the composites was analysed by scanning electron microscopy (SEM) images using a Hitachi S-2400. Particle lateral size was determined by laser diffraction in dry with a Mastersizer 2000.

Transmission electron microscopy (TEM) experiments were performed on a JEOL model JEM-2010 electron microscope.

Raman spectra were recorded on a confocal Renishaw inVia Raman microscope at room temperature. The system is equipped with a CCD detector and a holographic notch filter, using excitation wavelength of 532 nm. Scans were acquired from 1000 to 3400 cm$^{-1}$, performing maps of 25 spectra. Analysis and deconvolution of spectra were done in Wire 4.2 software. rGO-TPU samples were cryofractured and allocated on a glass substrate.

XPS analysis were performed is a Kratos Axis Ultra DLD.

BET was determined using and AUTOSORB-6 QUANTACHROME INSTRUMENTS. The samples were degassed in an AUTOSORB DEGASSER QUANTACHROME INSTRUMENTS at 250ºC for 8h.

Contact angle was determined using DSA method contact angle in a Kruss system in water in a pressed pellet sample.

Shore hardness was determined using a hardness tester HP-C from Bareiss (Germany). Virgin polymer was processed the same way as the composites to compare obtained data.

Electrical conductivity's variation with humidity was studied in a two-electrode cell using a Testo Huminator humidity chamber in a range from 10 %RH to 90 %RH and at 20 ºC, 30 ºC, and 40 ºC. Temperature and relative humidity was controlled with a Testo 350-XL.454 control unit to verify the values of the chamber. Electrical conductivity's variation with temperature was studied using a modified climatic chamber DYCOMENTAL CM1000 to maintain the RH constant at 10 %RH, humidity has been controlled with a Testo 350-XL.454. Humidity and climatic chamber and the humidity measurement system have been calibrated by ENSATEC, Spain.

Electrical conductivity was measured in an Autolab PGSTAT 302N in a frequency range from 1MHz to 1Hz and 350 mV amplitude to obtain electrochemical impedance spectroscopy (EIS) data.

**Table 1** Properties of rGO

| Reference | Apparent density ($g/cm^3$) | XY plane measured by laser diffraction LD50 and SEM ($\mu m$) | BET ($m^2/g$) | O content % by XPS | Contant angle |
|---|---|---|---|---|---|
| Flake Graphite | 0,640 | 0-500 | >3 | <0,2 | |
| rGO1 | 0,004 | 40±2 (20-25 SEM) | 289 | 2,63 | 85,4±4.4 |
| rGO2 | 0,002 | 39±2 (20-25 SEM) | 491 | 2,59 | 81,6±4,2 |
| rGO3 | 0,002 | 39±2 (20-25 SEM) | 467 | 0,90 | 95,3±1,6 |
| rGO4 | 0,002 | 42±2 (20-25 SEM) | 483 | 3,67 | 80,6±5,9 |
| rGO5 | 0,002 | 40±2 (20-25 SEM | 434 | 6,84 | 76,1±5,0 |
| rGO6 | 0,002 | 41±2 (20-25 SEM) | 511 | 11,52 | 70,3±1,5 |

## 3. Results and discussion

*3.1. Reduced Graphene Oxide characterization.*
High-yielding of 6 different rGO were obtained by a modified Hummers' method and a thermochemical reduction process.

In consequence, graphene related structures made of few layers of rGO with large lateral size (aprox. 40 µm), controllable BET and oxygen content can be produced for the preparation of Graphene Related Materials (GRM)-TPU composites.

The XY plane size was determined by SEM analyses and laser diffraction D50. Its lateral size has been reduced from 500 microns, of polycrystalline graphite particles, to approximately 40 µm size in the case of the rGO, determine by laser diffraction and 20-25 µm determined by SEM (Table 1). By SEM pictures, we can observe the discrete particle size and by laser diffraction we can also observe the size of the agglomerated particles. Oxidation and sonication processes may reduce the lateral size of rGO due to the break of the XY planes. [29]. In our case, no significant reduction of the lateral size, determined by laser diffraction or SEM, has been observed (Table 1).

Specific Surface Area (SSA) is another key characteristic of rGO. BET is related to the number of graphene layers and is a key factor for energy applications. The number of layers of rGO ($N_G$) could be calculated by dividing the maximum theoretical SSA of graphene by experimentally determined BET SSA ($N_G=2630/BET$) [30]. The BET SSA of prepared rGO (table 1) is far below the theoretical value of fully exfoliated pristine graphene. This is attributed to incomplete exfoliation during sonication and to the inaccessible surface caused by agglomeration [12, 31]. Based on the calculated $N_G$, all the rGO prepared can be categorized as multilayer reduced graphene oxide (rGO1: $N_G$~9) and few layers reduced graphene oxide (rGO2; rGO3; rGO4; rGO5; rGO6: $N_G$~5-6) based on Bianco's classification [32].

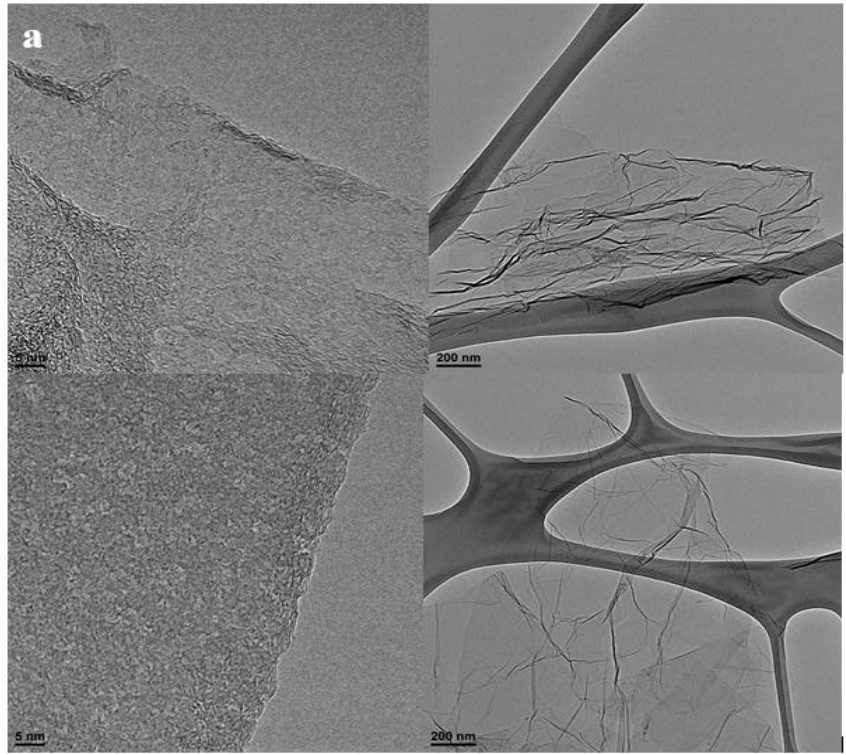

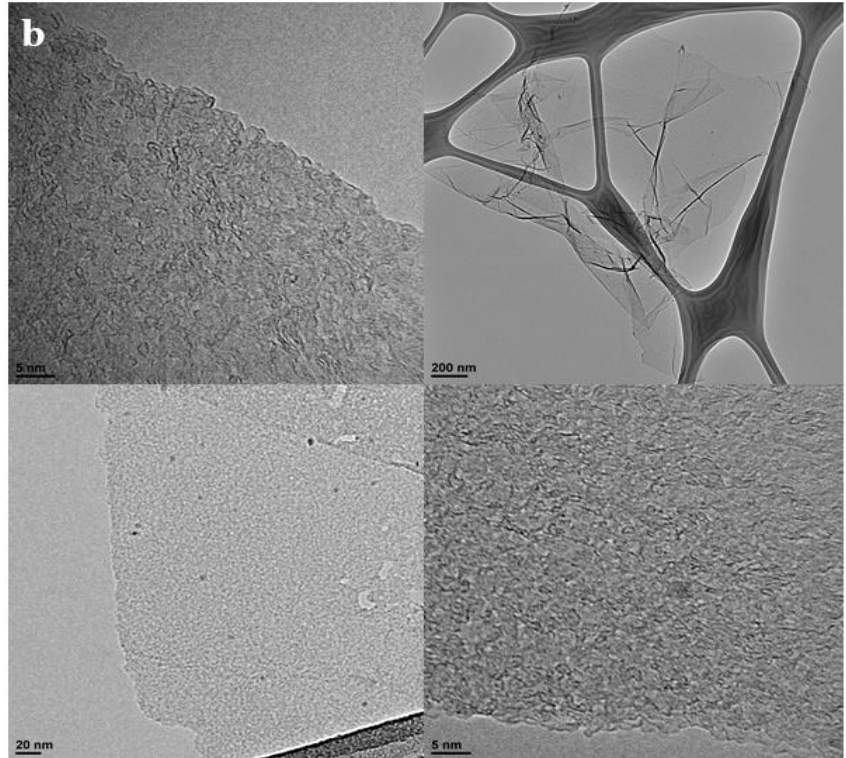

**Figure 1.** a) TEM image of rGO1 at different magnifications. b) TEM images of rGO2 at different magnifications.

TEM is used to qualitatively analyse the thickness of the rGOs prepared. Figure 1 shows TEM images of the rGO prepared. The rGO with high BET, rGO2, shows very low thickness in all particles in the TEM pictures, in most of them significantly lower than 1 nm. These TEM pictures are in well alignment with the other reference published results for rGO, see for example [33].

Raman spectra of the six rGO prepared are shown in figure 2. Raman spectra of the rGOs show the typical pattern for these materials: an intense D band presence ~1350 cm$^{-1}$ in all samples, which confirms the lattice distortions [34], a G$_{app}$ at ~1585 cm$^{-1}$ (1584 & 1585 cm$^{-1}$) corresponding to the superposition of 2 peaks the first-order scattering of the E$_{2g}$ mode of G and the contribution of D´ bands, 2D, D+D´ and 2D´ [35]. The increased I$_D$/I$_G$ ratio of rGO after the thermochemical reduction has been previously reported in literature [36] and the decrease in the I$_D$/I$_{Gapp}$ is also in good agreement with the transition from a more defective Stage 2 to stage 1.[37] However, as mentioned in [38], unreliability of the relationship between the I$_D$/I$_{Gapp}$ due to the overlap of G and D´ peaks, limits the utility of this relationship as a measure of density of defects in rGO, for that reason we have combine I$_D$/I$_{Gapp}$ vs Γ$_D$, (Figure 3a), correlating also with the XPS results. In figure 3b the relation between contact angle and $L_D$ obtained from Cançado´s formula[37] can be observed, showing and increase of the contact angle when increasing the distance between defects.

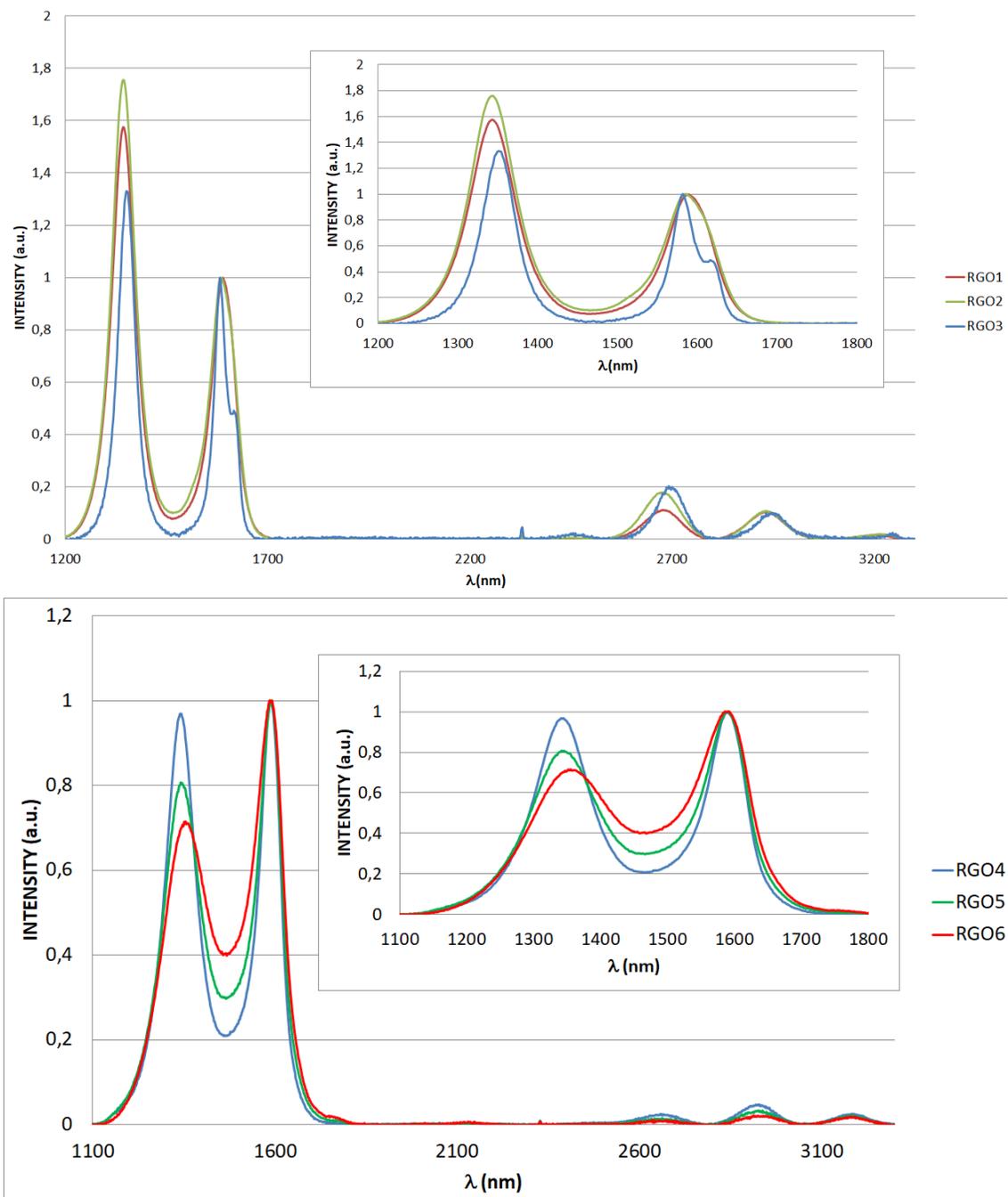

**Figure 2.** Raman spectra of a) rGO1, rGO2 and rGO3 b) rGO4, rGO5 and rGO6

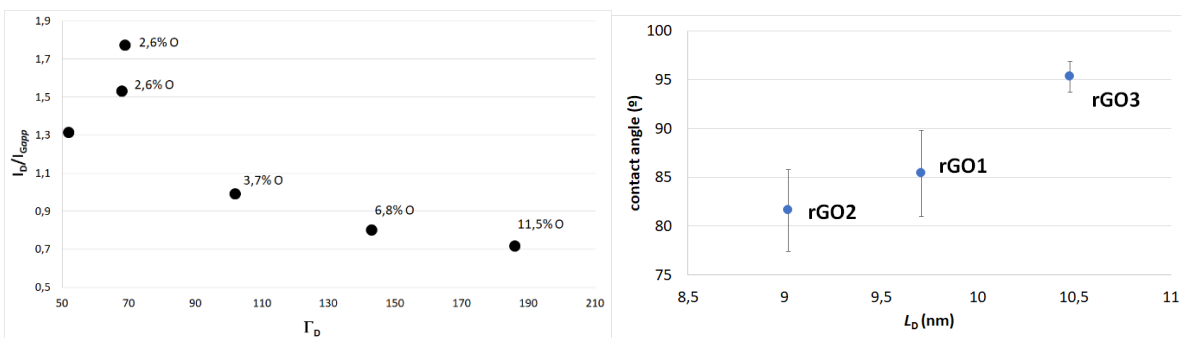

**Figure 3. a)** $I_D/I_{Gapp}$ vs $\Gamma_D$ for the rGO prepared and oxygen content determine from XPS. **b)** influence of distance between defects $\Gamma_D$ *vs* contact angle

Second order transition 2D´ has been used to calculate the inferred energy of D´ mode (D´$_{inf}$) and the differences between 2D´ (or D´$_{inf}$) and G$_{app}$. As previously reported, there is a shift to higher energies. The relation between the D´$_{inf}$ – G$_{app}$ and the C/O ratio obtained by XPS is also in good agreement with King *et al.* [38]. rGO1, rGO2 shows D´$_{inf}$ – G$_{app}$ = 4 for C/O 35:1; D´$_{inf}$ – G$_{app}$ = 5 and C/= 18:1 (rGO4); D´$_{inf}$ – G$_{app}$ = 6 and C/= 10:1 (rGO5), D´$_{inf}$ – G$_{app}$ = 7 and C/= 6:1 (rGO6). In the case of rGO3, the rGO with very low oxygen content, we have observed D´$_{inf}$ – G$_{app}$ = 43 for C/O 82:1.

XPS analysis show 2,59% of oxygen for the rGO1, 2,63% for the rGO2 and 0,90% for rGO3. The integration of the O 1s peak shows a component of C-O 58,2%, C=O 36,0% and COO 5,8% for the rGO1, C-O 64,1%, C=O 32,3% and COO 3,6% for the rGO2; in the case of rGO3, the low content of oxygen do not allow to do a correct integration of the peak. For the more defective rGO in Stage 2: C-O 48.04%, C=O 28.60% and COO 23,36 % for the rGO4, C-O 35,26%, C=O 36,87% and COO 27,87% for the rGO5 and C-O 35.11%, C=O 43.25% and COO 21,65 % for the rGO6.

*3.2. TPU nanocomposites characterization*

Morphologies of TPU nanocomposites were analysed by scanning electron microscopy: Representative SEM micrographs of TPU and the rGO are reported in figures 4, 5 and 6. In all composites, it is possible to observe regions with large aggregates of rGO particles with average lateral size of several tens to hundreds of microns aligned in the rGO XY plane, as well as other areas with a low content of rGO. At low magnification, higher number of particles can be observed for rGO2 compared with the rGO1 composites (see Figures 4,5,6 and Figure S1-S4 in Supplementary information) in agreement with the differences in BET values of the rGO. In the case of the composite with higher defects content rGO6, small improvement in the rGO-TPU interface has been observed probably due to the presence of COOH and OH groups and higher hydrophilic behaviour.

The average size of these aggregates increases with a larger amount of rGO in the composite. These aggregates create a conductive pathway consistent with the electrical properties observed for these composites. Similar behaviour has been observed for other TPU-GO composites previously published [39]. The thickness of these aggregates varies from few microns to 10 microns of stacking of parallel rGO nanosheets. At high loadings, higher stacking can be observed. This is also in agreement with electrical conductivity results observed for rGO1 and rGO2 composites. In the case of rGO6 composites, the size of the aggregates is higher compared with the other composites, this can be attributed to the higher hydrophilic character. Also the stacking of the rGO layer is more compacted that in the higher reduction composites.

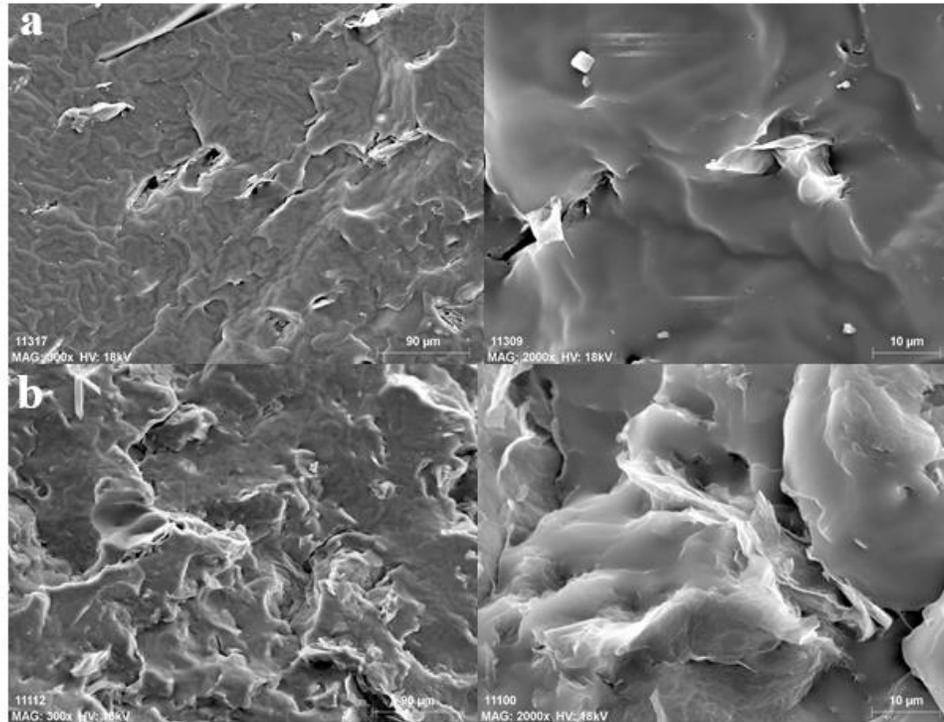

**Figure 4.** SEM pictures at different magnification of a) 0,25% rGO1; b) 1% rGO1

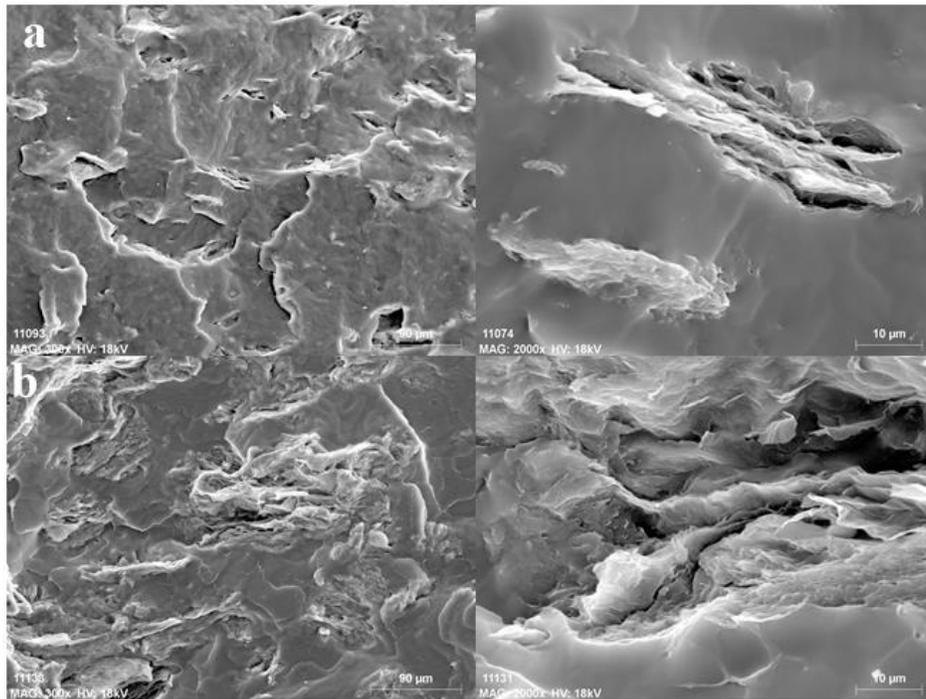

**Figure 5.** SEM picture at different magnification of a) 0,25% rGO2; b) 1% rGO2

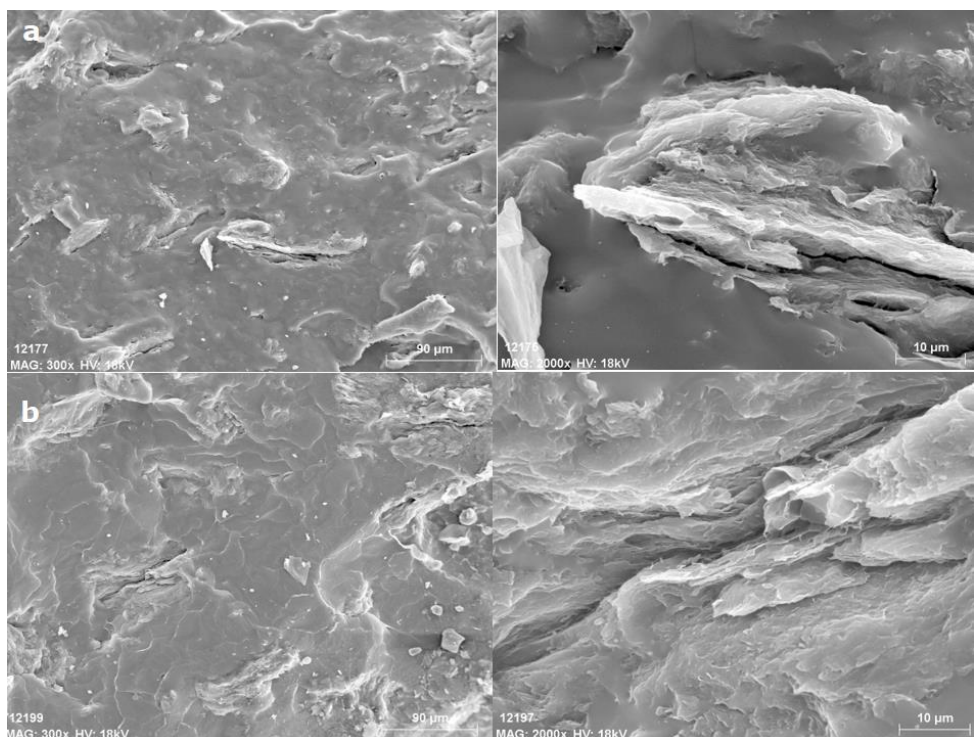
**Figure 6.** SEM picture at different magnification of a) 0,25% rGO6; b) 1% rGO6

At higher percentage of rGO, there is an increase in the size of the agglomerates, even higher than 100 μm. These agglomerates describe a pathway for the electron conduction consistent with the electrical conductivity observed for the composites described in this work.

Due to the overlap between the Raman spectra bands of rGO and TPU, we cannot discriminate their interaction in the rGO-matrix. (Figure S5 in Supplementary information)

The rGO used in the preparation of the nanocomposite has also influence over the hardness of the TPU composites (Figure 7). In all of the cases, there is an effective increase in hardness compared to neat TPU even at low loads of rGO, with an increase of hardness between 50% to 90% at 0,25 %wt. rGO2 to rGO6, the graphene materials with lower average number of layers, is more efficient in the increase of hardness than rGO1, especially at low loads. We have not observed any significant different on hardness of the rGO composites with high BET. At higher loading this difference is negligible, but hardness is double than the one obtained on the neat polymer. This fact is also consistent with the higher number of rGO particles.

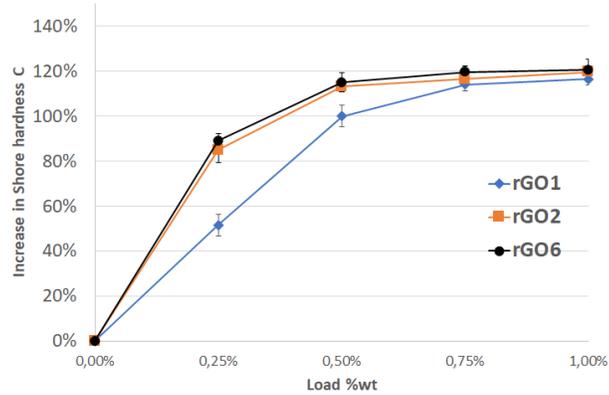

**Figure 7.** Hardness increase for rGO-TPU composites

*3.3 Electrical conductivity of the composites produced*

Electrical conductivity of rGO composites is highly dependent on the filler content, on the oxygen and defects content and on the degree of exfoliation of the rGO [13,14]. In this paper, we have modified the degree of exfoliation, defects and oxygen content and the number of particles keeping the  and lateral size of the rGO constant.

Electrical characterization has been carried out by alternating current techniques at 20 ºC and 40 %RH using complex impedance (Z) plot fittings and simulation of nyquist plots. Results of the electrical conductivity measurements of the composites, with different filler materials, are presented in Figure 8 showing electrical conductivity vs filler content of the rGO-TPU composites. rGO3 composite shows higher electrical conductivity, which can be attributed to the lower defects content that increase the intricsic conductivity of the rGO material. An increase of oxygen content produces a decrease in electrical conductivity and rGO5 and rGO6 do not exhibit ant conductivity. The electrical conductivity of rGO2 is higher than rGO1, this fact can be attributed to its ease of generating a conductive network due to the larger number of particles.

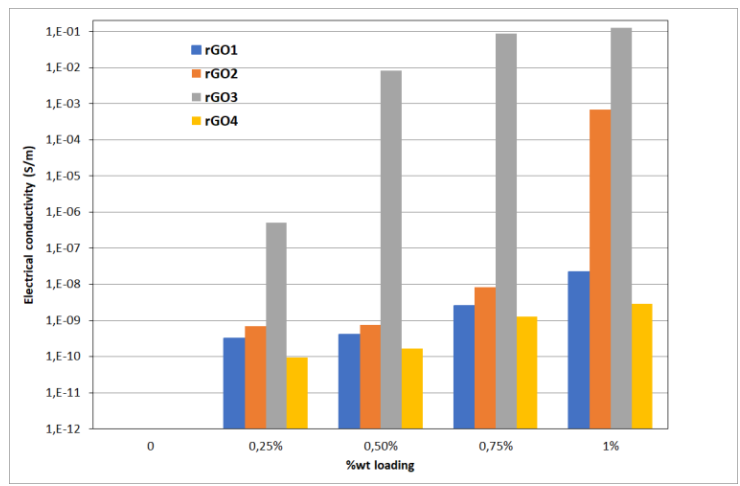

**Figure 8**. Electrical conductivity for rGO-TPU composites at 20 ºC and 40 %RH at different rGO content.

Figure 9 and S6 represent Nyquist plots, Z" (imaginary part of impedance) against Z' (real part of impedance), at some temperatures for the composite with 0,5% and 1% rGO2 in function of the temperature.

Tilted semi-circles are typical of carbon conductive fillers polymer composites (Figure S6) that fits to a simulated R(RQ) equivalent circuit. In figure 9 a second semicircle appears above 80 ºC for the 1 %wt rGO2 composite. The equivalent circuit used to fit the system was the same one from Figure S6 under 80 ºC and R(RQ)(RQ), displayed in figure 9, over this temperature, when the second semicircle appears.

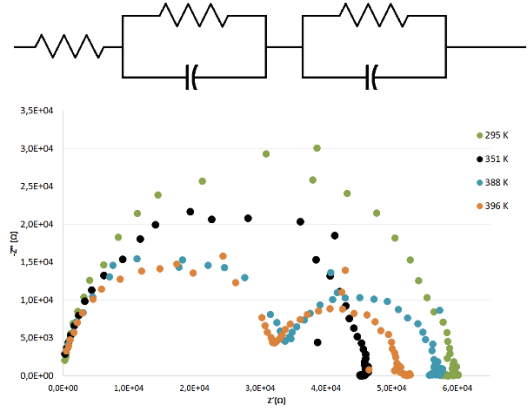

**Figure 9** Nyquist plots of the 1% rGO2 at different T.

To gain further knowledge, the effect of concentration and temperature on the electrical conductivity in the rGO-TPU composite, has been studied. Figure 10 shows the dc conductivity behaviour, which increases exponentially with temperature, indicating that the conductivity is thermally activated process. It can be expressed by the well-known Arrhenius equation (Equation 1).

$$\sigma = \sigma_0 e\left(\frac{-E_a}{k_B T}\right) \quad (1)$$

Where $k_B$ is Boltzmann constant. This is proportional to the number of charge carriers which can move, contributing to the electrical conductivity. The activation energy, $E_a$, obtained from the corresponding slope varies depending on the rGO used and the % of loading (Table 2). In all composites, a variation in the slope can be observed at a moderate temperature (Figure 10a). In the case of rGO2 composites a third slope over 100 ºC can be observed in the Arrhenius equation fitting for %wt below 1%.

These modifications in the slope can be attributed to the softening of the TPU composites and an increase in the freedom of movement of the polymer molecules, which allow the displacement of the rGO particles.

As it can be seen in table 2, there are not significant variations in the activation energy as the filler concentration increases. However, 1%wt rGO2 composite shows significant lower $E_a$ at all temperatures studied; this behaviour may be attributed to an increase of charge carrier density compared with the other fillers content. The experimental value of parameter $\sigma_0$ increases with increasing rGO2 content which confirms the increase of the charge carriers density. However, in rGO1 this effect is not clear.

**Table 2.** Activation energy in eV of the composites. T1: first change in the slope, T2: second change in the slope, σ₀

| Sample | Ea (Low T) | Ea (Medium T) | Ea (High T) | T1 & T2 | $\sigma_0$ |
|---|---|---|---|---|---|
| 0,25%-rGO1 | 0,12 | | 1,2 | 56 °C | 3,4E-08 |
| 0,50%-rGO1 | 0,46 | | 1,09 | 49 °C | 3,9E-06 |
| 0,75%-rGO1 | 0,28 | | 1,02 | 49 °C | 4,6E-07 |
| 1%-rGO1 | 0,31 | | 1,14 | 55 °C | 4,8E-07 |
| 0,25%-rGO2 | 0,1 | 0,75 | 1,48 | 40 °C, 82 °C | 6,4E-09 |
| 0,50%-rGO2 | 0,21 | 0,73 | 1,4 | 45 °C, 82 °C | 1,2E-07 |
| 0,75%-rGO2 | 0,33 | 1 | 3,81 | 53 °C, 103 °C | 2,5E-06 |
| 1%-rGO2 | 0,01 | 0,17 & 0,27 | | 50 °C | 7,3E-04 |

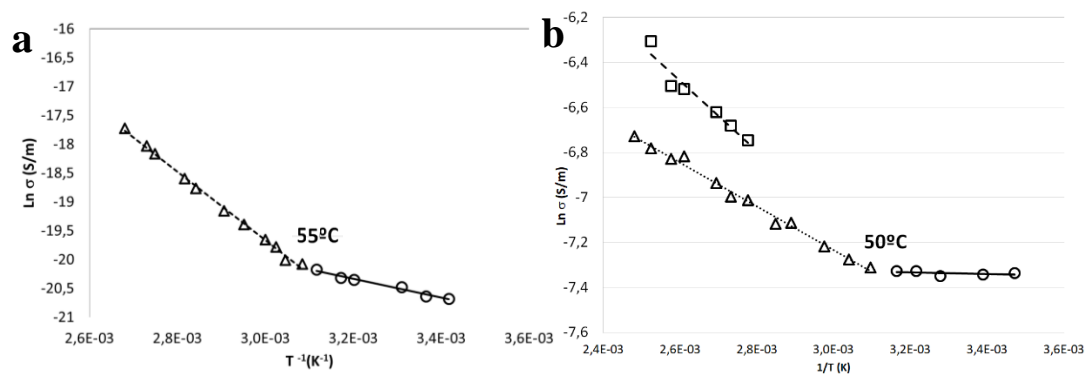

**Figure 10.** Electrical conductivity in function of the temperature for a)1%wt rGO1 b)1%wt rGO2 composite (O) fitting nyquist plot to one semicircle, fiiting to 2 semicircles: high frecuencies semicircle Δ; low frequencies semicircle □.

*3.4 Influence of humidity on the electrical properties of the composites.*

We have studied the influence of temperature and humidity in electrical conductivity of the rGO2-TPU and rGO3-TPU composite based on better electrical performance of these composites. rGO3-TPU composite has not shown any sensibility to the humidity (Figure S8). In Figure 11a the evolution of the electrical conductivity at particles loading below 1 %wt rGO2 (0,25 %wt to 0,75 %wt rGO2) can be observed. A significant increase of electrical conductivity over 40% with the increase of relative humidity (RH) can be observed for all the samples. In the case of the sample at 1 %wt lower influence of the humidity has been observed.

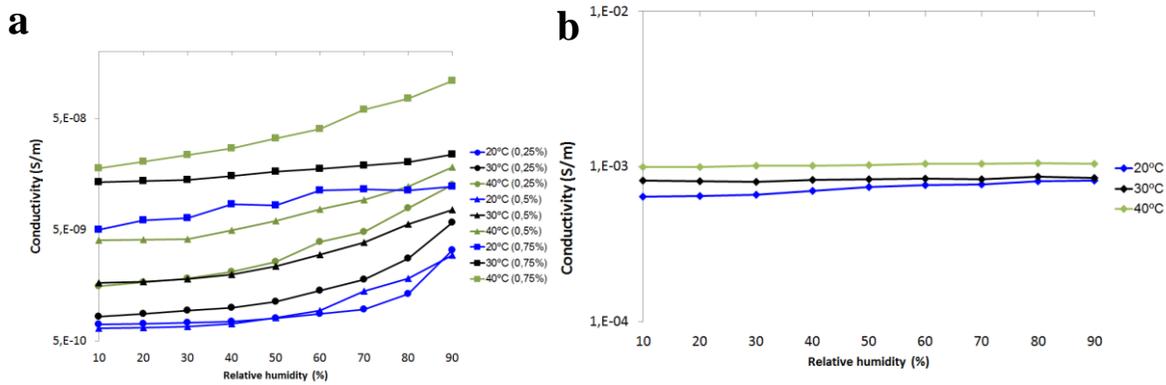

**Figure 11.** a) Electrical conductivity in function of %wt (0,25%; 0,5% and 0,75%) of rGO2, temperature and relative humidity. b) Electrical conductivity at 1%w rGO2 in function of the temperature and relative humidity

All the composites below 1 %wt show high sensitivity to the RH, higher than other carbon-based composites [41] and even better than the ones reported for graphene related materials composites [42], and in the same order of magnitude of some reported graphene/polyelectrolite composites [43]. Figure 12 shows the representation of the sensitivity of 2 rGO2 composites appling the formula $S = RH1/RH2$ [42], where RH1 and RH2 were the impedance at 10% RH and at RH respectively.

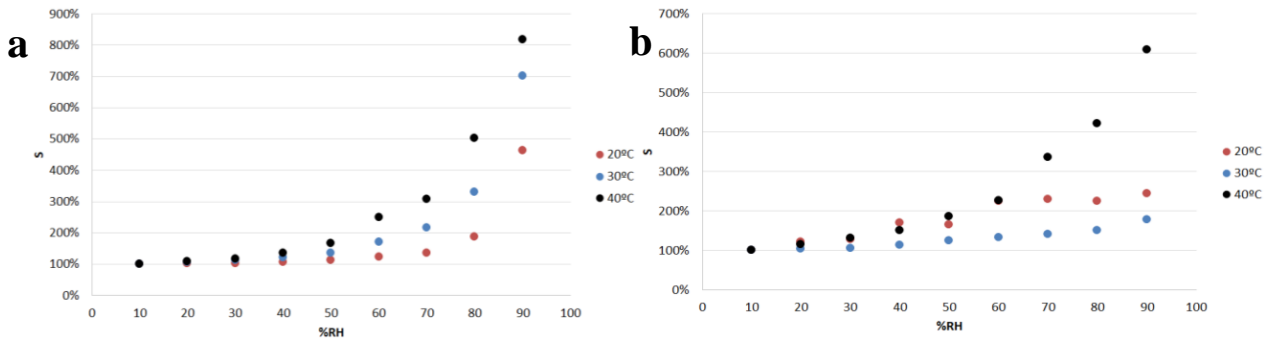

**Figure 12.** Variation of sensitivity at different temperatures with humidity at a) 0,25%rGO2 and b) 0,75% rGO2.

The response time of the composites (figure 13) is highly dependent on the electrical conductivity; for example, for 0,75 %wt rGO2 composite response time is about 2500 s. In the case of the composite at 1 %wt rGO2, the sensitivity is lower but the response time decreases to 160 s. These recovery times are higher than the observed for pure GO, that was reported a few tens of seconds [44]. However, in our case we are using a composite at just 1% of loading with good mechanical performance instead of the pure filler.

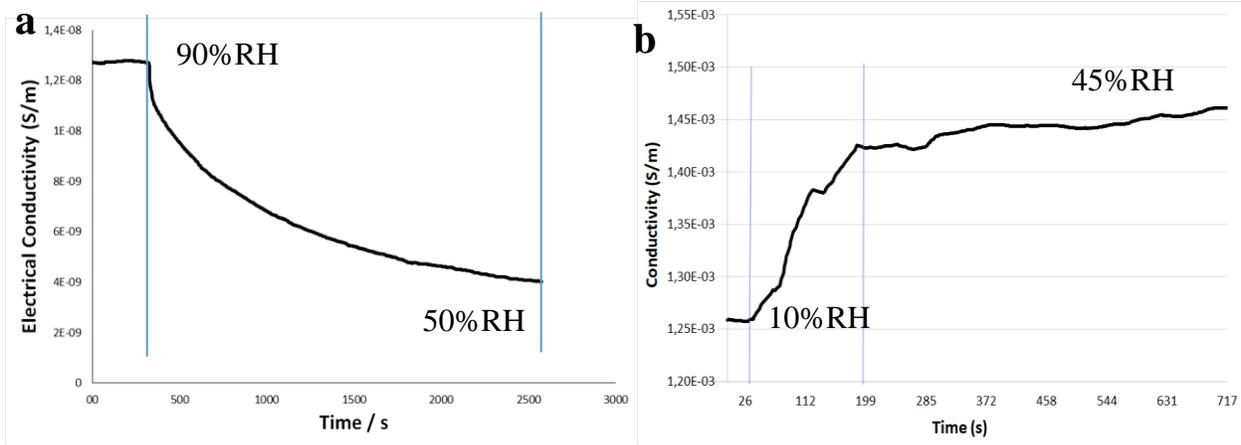

**Figure 13.** a) Variation of electrical conductivity in function of time from 90%RH to 50%RH for 0,75% rGO2 composite at 40ºC. b) Variation of electrical conductivity in function of time from 10 %RH to 45 %RH for 1% rGO2 composite at 40ºC.

Figure S7 represents Nyquist's plots for the composite with 0,75% rGO2 in function of the RH. As already mentioned, tilted semi-circles are typical of carbon conductive fillers polymer composites that fits to a simulated R(RQ) equivalent circuit. As it can be observed, the impedance decreases as RH rises, similar behaviour has previously been reported [43].

*3.5 Dielectric relaxation*

By EIS we have obtained and represented conductivity of nanocomposites at various temperatures in a range of frequencies between 1MHz to 1Hz.

It is possible to observe well-differentiated regions in Figure 14; at high and low frequencies. Low frequencies plateau regions correspond to the frequency independence on conductivity, called $\sigma_0$, and it is calculated by extrapolation. According to Jonscher´s universal law or UPL [18] (equation 2), we have analysed the dispersion region, calculating the parameters summed in table 2.

$$\sigma = \sigma_0 + A\omega^n \qquad (2)$$

Where $A$ is a pre-exponential constant, $\omega$ is the angular frequency and $n$ is the power law exponent, which usually in literature is stated to have values between 0 and 1 [45]. It represents the degree of interaction between the rGO and the polymer matrix.

In prepared rGO-TPU nanocomposites, power law exponent, or n-factor, is highly dependent on the percentage of filler. For low concentration of rGO2 (≤0,75%wt) and in all the rGO1 composites at all the temperatures studied, n factor is higher than the unit; which is an incongruence according to UPL [46, 47]. This behaviour has been also observed in some ionic and electronic conductors [26-28, 48, 49] and underlined by Papathanassiou et al in conductive polymers [50] and disordered material [51, 52]. We have also observed n-factor dependence on frequency in some of the composites at room temperature, which is also unjustified in the UPL [50] (see for example figures 14a and b). Other references of n-factor for graphene materials and CNTs composites present values in accordance with Jonscher´s Universal Law (between 0,7-0,9 at room temperature) [53-55]. At higher loading, the n-factor decreases.

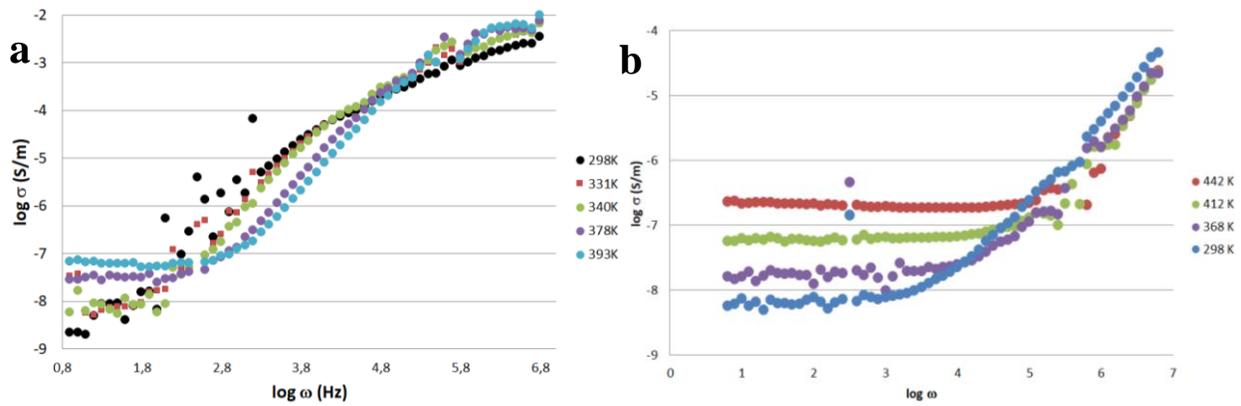

**Figure 14**. Frequency dependence of electrical conductivity at various temperatures a) 1%wt rGO1-TPU. b) 0,5%w rGO2-TPU.

The frequency dependence of electrical conductivity and *n* is the power law exponent was analysed by the logarithm of σ as a function of ω for different temperature and RH. In the case of 0,75% rGO2, n-factor is still over 1 composite at 40ºC, we can no observe differences in n factor with RH. Figures 15 and S9 shows an example of the comparison of the influence of humidity and temperature in the frequency dependence of the electrical conductivity for rGO2 at 0,75%w. The influence of the RH is more remarkable than the temperature effect in the composite.

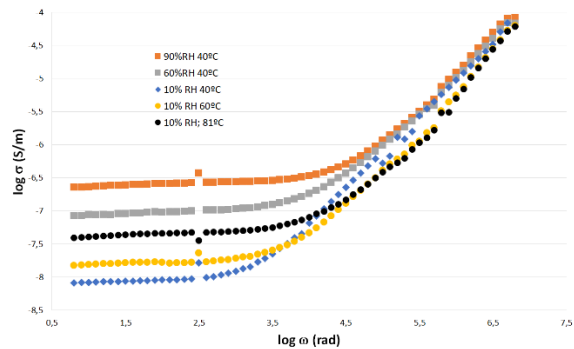

**Figure 15.** Frequency dependence of electrical conductivity of 0,75%w rGO2-TPU nanocomposite at different %RH and temperatures.

Mauritz and Papathanassiou [19, 52] reported the insensitivity of -logA/n to material composition, structural transition and temperature. For the composite rGO2/TPU at loading of 0,5% and 0,75% we have observed that the -logA/n ratio is almost constant with temperature, however, in the 1%rGO2 composite, there is linear evolution with temperature (Figure 16).

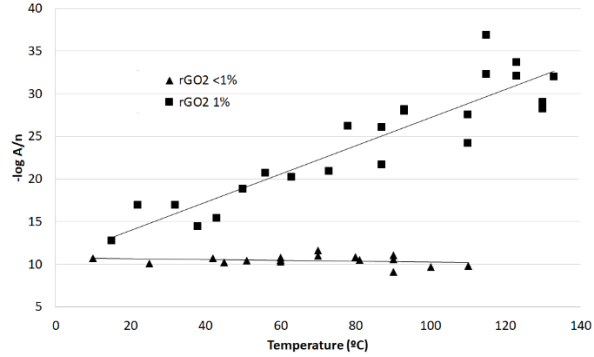

**Figure 16** -logA/n vs temperature for square rGO2 at 1%wt, triangle 0,5 and 0,75% rGO2.

Figure 17 and S10 illustrates -logA *vs* n factor. Log A is proportional to the temperature and to RH evolution of the n-factor in the 0,75% rGO2 composite (Figure S10a). In both cases, linearity is observed all along the range of temperatures and %RH.

The -logA *vs* n plot in figure S10b illustrates the proportional evolution of log A which is proportional to the temperature evolution of the n-factor in the 1% rGO2 composite. In the case of the second domain, a small deviation from the linearity has been observed.

It is important to remark that all for the experimental results based on rGO2 composites there is a proportional evolution of log A *vs* n-factor (Figure 17, a straight line $R^2 = 0,9984$ has been fitted to the data points) at different RH, temperatures and even rGO loading, however it is different depending on the rGO material used. In the case of rGO4 similar behaviour that rGO1 has been observed and the difference in behaviour is negligible. Similar behaviour, based on temperature evolution has been observed in CNTs/epoxy systems [19]. Figure S11 illustrates -logA *vs* n factor and the differences on the graphene material and matrix values extracted from [53, 54] for different graphene materials/matrix systems.

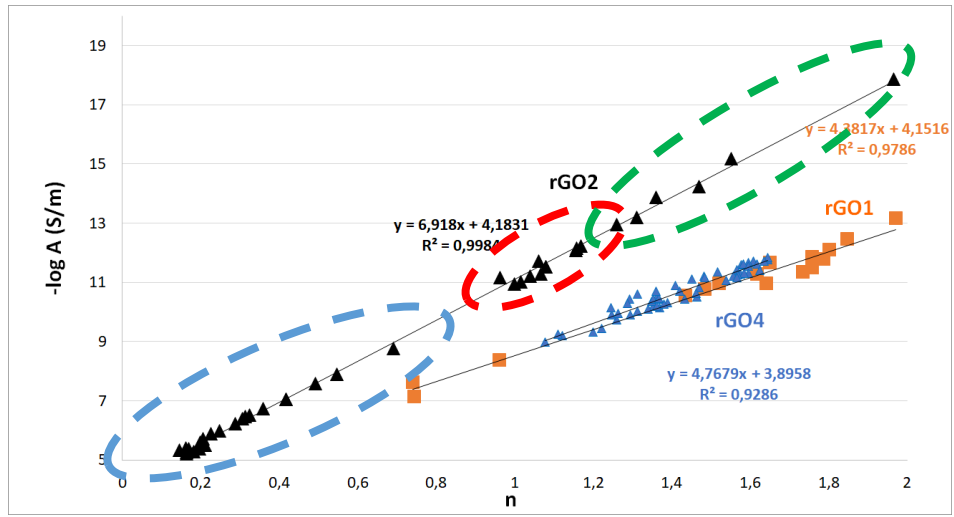

**Figure 17** – Black triangles: log A vs n for rGO2/TPU composites at different temperatures and humidity. Blue ellipse at 1%wt; red at 0,75% and green 0,5%. Orange squares are for rGO1 composites and blue triangles for rGO4.

To study the transport mechanisms in our rGO composites we have used the Mott equation (3) [56],

$$\sigma = \sigma_0 \exp\left[-\left(\frac{T_{Mott}}{T}\right)^{\gamma}\right] \qquad (3)$$

The variable range hopping (VRH) exponent γ determines the dimensionality (d) by the relation γ = 1/1+d. The possible values of γ are 1/4, 1/3, 1/2 for three-dimensional, two-dimensional and one-dimensional systems, respectively.

We have studied the variation of the electrical conductivity and represent the different models to determine the exponent γ. On the range of temperatures studied, there is not an unequivocal behaviour (See Figure S12-S19 in supporting information), however for all the composites the maximum adjustment for γ is 1/4 (Figure 18), which is in accordance with other CNT [57] and graphite composites reported results [58, 59]. In the case of disorder deposited graphene materials [60, 61] it follows a 2D system behaviour.

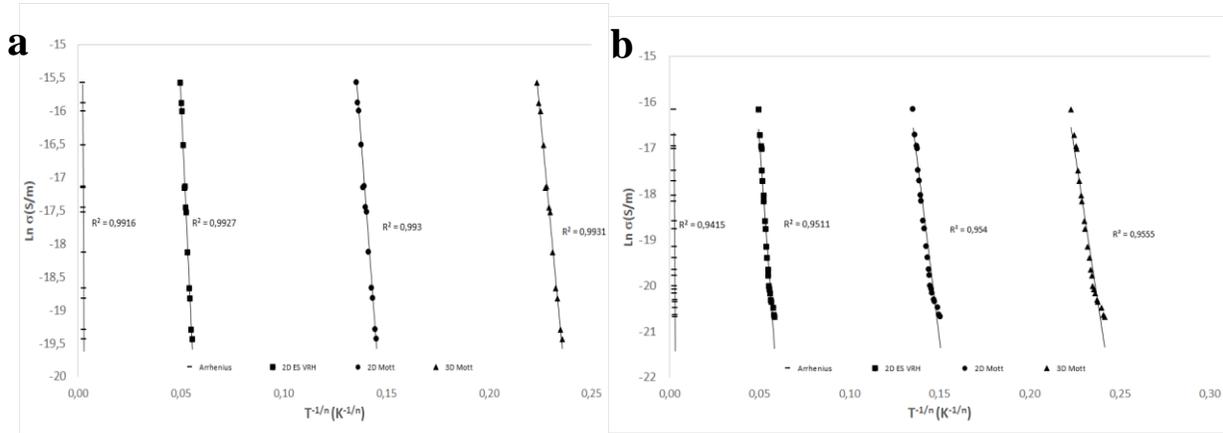

**Figure 18.** Fitting model for rGO1: a) 0,25%w b) 1%w

We have observed that relaxation time, τ, decreases (calculated from ω$_p$ in the Z´´ plots) with an increasing on temperature and RH (figure 19). In the case of RH, there is a linear decay and in the case of the temperature, an exponential decay that follows Arrhenius behaviour (Equation 4) is observed. The relaxation time decreases when increasing the filler content (figure 19 inset) similar to MWCNTs [62-64]. The difference in the E$_a$ of the relaxation time (0,39 and 0,11 eV for 0,75% and 1% rGO2 respectively) is indicative of the higher thermal influence in the lower loading reduce graphene oxide composite which indicates a faster movement of mobile electron in the 1% loading composite.

$$\tau = \tau_0 e^{\left(\frac{-E_a}{k_B T}\right)} \qquad (4)$$

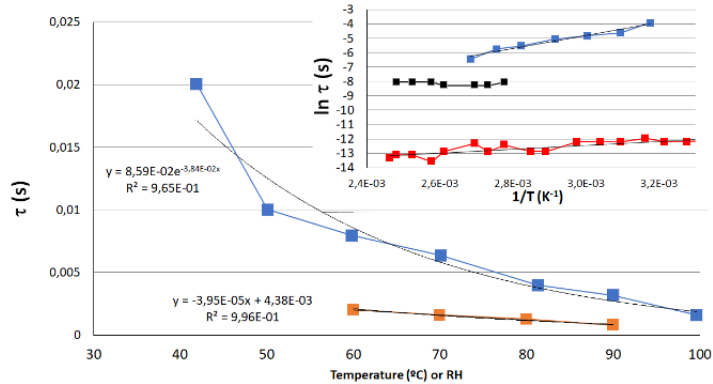

**Figure 19**. τ vs temperature (blue lines) and humidity (orange line) for 0,75% rGO2 composite. Inset ln τ vs 1/T (blue line, 0,75%rGO2, red line 1%rGO2 first component, and black line second time domain.

## 4. Conclusions

We have prepared tuneable in thickness and defects and oxygen content reduced graphene oxide using a modified Hummers' method, as it can be seen by TEM, SEM and Raman measurements. Using solution blending, we have produced TPU composites with the rGO prepared. We have observed high influence on BET in the electrical conductivity of the rGO-TPU composites, which can be related to the average thickness.

We have determined the influence of humidity and temperature on the electrical conductivity of rGO-TPU systems, showing high sensibility and low response time to humidity.

On one hand, depending on the filler content, temperature and humidity, a frequency dependence on electrical conductivity can be experimentally determined. On the other hand, at high loadings, temperature and humidity; a lower dependence on the frequency has been observed.

There is not an unequivocal behaviour in the determination of the transport mechanism; however, for all the composites the maximum adjustment for γ is 1/4 which is in good agreement with a 3D-VRH mechanism, similar to the one observed in other graphene materials or CNTs polymer composites.

According to UPL, we have shown different n factors from 0,1 to 2 depending on the filler content, type of filler, temperature and humidity conditions on the rGO-TPU composites. Moreover, it was determined that -logA *vs* n follows a linear behaviour even at different loadings of rGO, temperatures and humidity, nevertheless, there is a different behaviour depending on the rGO used for the preparation of the composite.

**Authors contributions**

J. Gomez and E. Villaro conceived the experiments and coordinated the project and all the data analysis, E. Villaro carried out the entire preparation of nanocomposites, A Navas contributed to the electron microscopy characterization, I. Recio contributed to the electrical characterization. Manuscript was written by J. Gomez and E. Villaro.

**Acknowledgments**

A Navas, E Villaro and J. Gomez thank the funding from European Union H2020 Programme under grant agreement n°696656 Graphene Flagship Core1. I Recio thanks the funding from European Union H2020 Programme under grant agreement n°642890 The Link.